\documentclass[letterpaper]{article}

\usepackage{prepr}

\title{SAFER-D: A Self-Adaptive Security Framework for Distributed Computing\\Architectures}
\author{
\PREPauthor{Marco Stadler}{Johannes Kepler University Linz, LIT Secure and Correct Systems Lab / Institute of Business Informatics -- Software Engineering}{marco.stadler@jku.at}
\PREPauthor{Michael Vierhauser}{University of Innsbruck, Department of Computer Science}{michael.vierhauser@uibk.ac.at}
\PREPauthor{Michael Riegler}{ENGEL Austria GmbH, Information Security}{michael.riegler@engel.at}
\PREPauthor{Daniel Waghubinger}{Johannes Kepler University Linz}{N/A}
\PREPauthor{Johannes Sametinger}{Johannes Kepler University Linz, LIT Secure and Correct Systems Lab / Institute of Business Informatics -- Software Engineering}{johannes.sametinger@jku.at}
\vspace{0.6cm}
}
\setvenue{the \vspace{3pt}\textbf{\\19th European Conference on Software Architecture (ECSA)}}
\setvenueone{the \textbf{19th European Conference on Software Architecture (ECSA)}}
\setyear{2025}
\setdoi{10.1007/978-3-032-02138-0\_13}

\usepackage{harveyballs}

\newcommand{\fwn}{SAFER-D\xspace}
\newcommand{\fwcomp}[1]{\ttformat{#1\xspace}}

\definecolor{myblue}{HTML}{007FFF}
\definecolor{myorange}{HTML}{FF8000}
\definecolor{oliveish}{HTML}{ccbb44}
\definecolor{pinkish}{HTML}{ee6677}
\definecolor{greyish}{HTML}{bbbbbb}
\definecolor{blueish}{HTML}{66ccee}
\definecolor{greenish}{HTML}{6ab797}

\newcommand{\fwblue}[1]{{\color{myblue}{#1}}}
\newcommand{\fworange}[1]{{\color{myorange}{#1}}}

\newcommand{\mapecharlocal}[1]{\tikz[baseline=-0.5ex]\node[draw, scale=0.8, fill=myblue, text=white, circle, inner sep=0, minimum size=1.25em] {\fwcomp{#1}};}
\newcommand{\mapecharglobal}[1]{\tikz[baseline=-0.5ex]\node[draw, scale=0.8, fill=myorange, circle, inner sep=0,  minimum size=1.25em] {\fwcomp{#1}};}
\newcommand{\mapecharapi}[1]{\tikz[baseline=-0.5ex]\node[draw, scale=0.8, fill=greenish, circle, inner sep=0,  minimum size=1.25em] {\fwcomp{#1}};}

\newcommand{\ttformat}[1]{{\texttt{\small #1}}}

\definecolor{alizarin}{rgb}{0.82, 0.1, 0.26}

\newcommand{\ucitem}[1]{
\noindent$\bullet$ \textbf{UC-#1 -- }
\ifthenelse{\equal{#1}{1}}{\textbf{Edge Computing:}\xspace}{\textbf{Multi-Modal Auth. System:}\xspace}%
}

\colorlet{punct}{red!60!black}
\definecolor{background}{HTML}{EEEEEE}
\definecolor{delim}{RGB}{20,105,176}
\colorlet{numb}{magenta!60!black}
\lstdefinelanguage{json}{
    basicstyle=\ttfamily\footnotesize,
    numbers=left,
    numberstyle=\scriptsize,
    stepnumber=1,
    numbersep=8pt,
    showstringspaces=false,
    breaklines=true,
    frame=lines,
    literate=
     *{0}{{{\color{numb}0}}}{1}
      {1}{{{\color{numb}1}}}{1}
      {2}{{{\color{numb}2}}}{1}
      {4}{{{\color{numb}4}}}{1}
      {5}{{{\color{numb}5}}}{1}
      {6}{{{\color{numb}6}}}{1}
      {7}{{{\color{numb}7}}}{1}
      {8}{{{\color{numb}8}}}{1}
      {9}{{{\color{numb}9}}}{1}
      {:}{{{\color{punct}{:}}}}{1}
      {,}{{{\color{punct}{,}}}}{1}
      {\{}{{{\color{delim}{\{}}}}{1}
      {\}}{{{\color{delim}{\}}}}}{1}
      {[}{{{\color{delim}{[}}}}{1}
      {true}{{{\color{numb}{true}}}}{1}
      {]}{{{\color{delim}{]}}}}{1},
}

\newcounter{rqcounter}
\newcommand{\RQ}[1]{%
    \refstepcounter{rqcounter}%
    \noindent\textbf{\hypertarget{rq:\therqcounter}{RQ\therqcounter: }}\textit{#1}%
    \label{rq:\therqcounter}%
}
\newcommand{\RQref}[1]{\hyperlink{rq:#1}{RQ#1}}

\makeatletter
\patchcmd{\thebibliography}
  {\section*{\refname}}%
  {\section*{\refname}}
  {}{}
\makeatother

\definecolor{reviewbg}{RGB}{188, 222, 255} 

\newcommand{\hball}[1]{\raisebox{-0.15em}{\csname harveyBall#1\endcsname}}

\DeclareCaptionLabelFormat{purplelabel}{\textcolor{purple}{#1~#2:}}

\usepackage[most]{tcolorbox}

\begin{document}

\maketitle

\begin{abstract}
    
The rise of the Internet of Things and Cyber-Physical Systems has introduced new challenges on ensuring secure and robust communication.
The growing number of connected devices increases network complexity, leading to higher latency and traffic. 
Distributed computing architectures (DCAs) have gained prominence to address these issues. 
This shift has significantly expanded the attack surface, requiring additional security measures to protect all components -- from sensors and actuators to edge nodes and central servers. 
Recent incidents highlight the difficulty of this task: Cyberattacks, like distributed denial of service attacks, continue to pose severe threats and cause substantial damage.

Implementing a holistic defense mechanism remains an open challenge, particularly against attacks that demand both enhanced resilience and rapid response.
Addressing this gap requires innovative solutions to enhance the security of DCAs.

In this work, we present our holistic self-adaptive security framework which combines different adaptation strategies to create comprehensive and efficient defense mechanisms.
We describe how to incorporate the framework into a real-world use case scenario and further evaluate its applicability and efficiency. Our evaluation yields promising results, indicating great potential to further extend the research on our framework.

\end{abstract}

\section{Introduction}
\label{sec:intro}


The hype surrounding the Internet of Things (IoT) and Cyber-Physical Systems (CPSs)  drives a surge in Internet-connected devices.
Managing this many devices requires careful organization, often achieved through diverse architectural styles and patterns~\cite{el-sayed_edge_2018}. 
Instead of a centralized infrastructure, distributed computing architectures (DCAs) are used to provide reduced latency, real-time analysis, high scalability, low operational cost, and improved quality of service~\cite{el-sayed_edge_2018}.
Although distributed computing helps operators deal with complexity, it affects the attack surface of these architectures.
The sheer number of devices and their convolution, heterogeneity, diversity, interoperability, portability, mobility, location, topology, and distribution of objects cause an increase in attack surface and make the architecture susceptible to cyberattacks (details in ~\cite{sadhuInternetThingsSecurity2022}).
In particular, interoperability and interdependency are crucial factors in this context~\cite{momohSmartGridFundamentals2012}. 
A failure caused by an attacker to one subsystem can lead to cascading failures, rendering the whole DCA inoperable~\cite{buldyrevCatastrophicCascadeFailures2010}. 
Particularly on critical infrastructure, a successful cyberattack can cause severe harm~\cite{thakur2016impact,palletiCascadingEffectsCyberattacks2021, carloCyberAttacksCritical2024}, ranging from compromised databases to human injury.
Recent incidents, such as the record-breaking 5.6 Tbps Distributed Denial of Service (DDoS) attack targeting Cloudflare's infrastructure~\cite{Recordbreaking56Tbps2025}, highlight the tangible risks faced by the industry. 
Studies~\cite{thakur2016impact,
palletiCascadingEffectsCyberattacks2021,carloCyberAttacksCritical2024} demonstrate that the scientific community recognizes these threats, emphasizing the urgent need for resilient security strategies in DCAs.


The reasons for such incidents still occurring are manifold. 
Security measures often only provide isolated and passive defense mechanisms, severely limiting their effectiveness~\cite{xiaoEdgeComputingSecurity2019}. 
Passive mechanisms usually rely on predefined rules.
For example, they may block network packets based on known signatures~\cite{otoumASIDSAnomalySignature2021}.
They follow a ``detect then patch'' philosophy, meaning they are only effective after an attack.
As a result, they cannot adapt proactively and respond to threats in real-time~\cite{xiaoEdgeComputingSecurity2019}. --
As a result, the question of securing DCAs with a holistic and active security solution to efficiently adapt to the evolving threat landscape remains.
In this paper, we propose a novel idea of combining self-adaptive architectural patterns to improve the security of DCAs.
More specifically, we leverage, among others, hierarchical adaptation strategies~\cite{weynsPatternsDecentralizedControl2013}, adaptation strategies used in Systems of Systems (SoS)~\cite{weynsChallengesSelfadaptationSystems2013},  and the concept of security levels (cf.~\citesec{sec-levels}) to enhance the overall resilience of DCAs in the event of attacks.
To the best of our knowledge, no prior work combines hierarchical, collaborative, and decentralized adaptation strategies to ensure self-adaptive security under partial system failure. Our framework addresses this gap through its dual-loop architecture and adaptation modes.
As part of this, we claim the following contributions:%
\begin{itemize}[leftmargin=*, topsep=0pt, parsep=0pt] 
    \item Novel, \underline{S}ecure \underline{A}daptive \underline{F}ramework for \underline{E}fficient \underline{R}esilience in \underline{D}istributed computing architectures (\fwn) that allows for security adaptations at the architectural level, even when under attack.
    \item Prototypical implementations using real-world edge computing architectures for component reuse. 
    \item Evaluation of applicability and efficiency of \fwn, based on realistic security use case scenarios.
\end{itemize}

\section{Motivating Architectural Challenges}
\label{sec:background}

In an increasingly interconnected world, the ability to dynamically adapt to emerging threat scenarios is becoming ever more critical.
Adaptive threat monitoring focuses on continuously observing systems for unusual or suspicious activities and adjusting responses based on evolving contexts. 
This adaptability is vital in defending against cyberattacks targeting CPSs/IoT systems where static approaches are insufficient.
Intrusion Detection Systems (IDS)~\cite{liaoIntrusionDetectionSystem2013}  often leverage an adaptive approach to detect and withstand cyberattacks.
An IDS commonly monitors network or system activities, detects potential security threats, and executes appropriate countermeasures or sends alerts. -- 
This process largely aligns with the MAPE-K (Monitor, Analyze, Plan, Execute on a shared Knowledge base) loop~\cite{kephartVisionAutonomicComputing2003}, a foundational pattern for self-adaptive system architectures.

Consider the following DCA example, which will serve as our running case: Edge computing, combined with fog and cloud computing, places substantial computing and storage resources at the (physical) outer ``edges,'' where data is generated.
The system processes data directly, forwards only aggregated data to fog computing components, aggregates it again, and then sends only the relevant data to the next central server, continuing this pattern\cite{caoOverviewEdgeComputing2020}.
An IDS applied to one of the edge devices of such an edge computing architecture  \textit{monitors}, for instance, the network traffic on the device, \textit{analyzes} the data to identify anomalies or suspicious patterns (e.g., a DDoS flooding attack), \textit{plans} appropriate responses (e.g., block a specific Internet Protocol (IP) address), and \textit{executes} the countermeasures (append IPs to a blocklist) or generates alerts, all supported by a \textit{knowledge} base to enhance detection accuracy and adaptability.
In the following, we highlight \textbf{Architectural Challenges (ACs)} concerning a self-adaptive security framework for DCAs.
These challenges are informed by our industry collaborations and supported by academic literature, based on recurring needs identified in regular technical meetings and through a structured review of recent research on self-protective and adaptive systems.

\textbf{AC~1 - Managing adaptation in complex and large-scale DCAs using a single MAPE-K loop is insufficient:}
Weyns~\etal formalized a series of architectural patterns comprising multiple interconnected MAPE-K loops to deal with large, complex, and heterogeneous systems~\cite{weynsPatternsDecentralizedControl2013}.
Among others, they introduced the hierarchical control pattern.
This pattern manages the complexity of self-adaptation by establishing a layered separation of concerns through a hierarchy of MAPE-K loops.
Loops at lower layers operate on a short time scale, ensuring that the portion of the system under their direct control adapt promptly. 
Higher levels operate on a more global/strategic scale over an extended period. Ultimately, the MAPE-K loop at the system's summit determines the system's overarching adaptation objectives.
Applying the hierarchical MAPE-K pattern to our aforementioned example implies that, for instance, the fog nodes in the edge computing architecture use the monitoring data of multiple underlying edge devices for the adaptation loops and then roll out a collective adaptation strategy for all devices associated with the respective fog node. 

\textbf{AC~2 - Hierarchical adaptation strategies break when intermediate nodes are compromised:}
In our example, the edge computing architecture follows a hierarchical organization, which creates dependencies.
Suppose fog node in the hierarchy is compromised and unavailable, e.g., due to a successful attack.
In that case, the underlying edge devices will not receive adaptation updates and thus remain susceptible to subsequent cyberattacks.
Regarding security, individual system components must be independent from an operational and managerial viewpoint, exhibiting SoS characteristics~\cite{weynsChallengesSelfadaptationSystems2013, chambersSelfAdaptationLooselyCoupled2024}.
Self-adaptive architectural patterns for SoS have been frequently studied~\cite{weynsChallengesSelfadaptationSystems2013}.
One of them, the \textit{Collaborative Adaptations} style, allows for adaptations on a control-theoretic level.
Additionally, the architecture allows for interactions among the managed systems, supporting collaboration between the subsystems. Using this pattern, the SoS can adapt comprehensively while considering each component~\cite{weynsChallengesSelfadaptationSystems2013}.
Subsystems can then coordinate locally and continue adaptation without relying on the compromised node.

\textbf{AC~3 - Security mechanisms must remain effective even when parts of the system are already compromised:}
{The Risk of an attacker compromising a system is defined as follows~\cite{kaplanQuantitativeDefinitionRisk1981}:

\begin{equation}
    R = \{s_{i}, p_{i}, x_{i}\},\quad \quad i=1, 2, ..., N
    \label{eq:risk}
\end{equation}
where  $R$ represents the risk; $s$ an undesirable event scenario description;  $p$ the probability of the scenario; $x$ the potential damage caused by the scenario; and $N$  the number of possible scenarios that may cause damage to a system. It is important to note that $p$ and $x$ are not constants but rather evolve over the time of an attack, serving here as a conceptual model to illustrate this dynamic.
In our edge computing example, if an attacker compromises one edge device, the probability that they successfully compromise another edge device increases (i.e., $p$ in Eq.~\eqref{eq:risk} evolves).
If attackers can circumvent security mechanisms once, they can reproduce the attack on other devices with the same security mechanisms. Similarly, once a publicly available proof of concept exploit exists for a known vulnerability, the probability of reproduction increases.
A compromised single system can also influence the ``neighbors'' communicating with the device. Depending on the type of attack, the attackers move laterally~\cite{heComprehensiveDetectionMethod2024} and propagate~\cite{acaraliModellingSmartGrid2022} to the controlling (managing) systems~\cite{larkinEvaluationSecuritySolutions2014}, i.e., in the best case (from an attacker's perspective), up the hierarchy to attain more and more control.
Security solutions are required to respond promptly to attacks and must be able to cope with already compromised system components.

\textbf{AC~4 - Security adaptations must consider the evolving criticality and impact of threats over time:} Not only does the probability evolve, but also the potential damage (cf.~$x$ in Eq.~\eqref{eq:risk}). 
For instance, if the running example’s edge computing system supports autonomous driving, it must prioritize safety to protect human lives~\cite{liuEdgeComputingAutonomous2019}.
A collision becomes more likely if the vehicle malfunctions (due to an attack).
Autonomous vehicles must slow down, or even shut down entirely, after detecting malicious activity.
To cope with these evolving factors, security mechanisms must support means of criticality.
\section{The \fwn Framework}
\label{sec:approach}

In the following, our novel \fwn framework addresses the architectural challenges (\textbf{AC~1-4}). 

\subsection{Core Components}
\label{sec:core-components}
\citefig{high-level-fw} depicts a high-level overview of \fwn.
The core idea is that \fwn is deployed on every single subsystem of the DCA. 
In the running case, this implies server, fog, and edge computing subsystems each run an instance of \fwn. 
Naturally, these instances must be tailored to the respective hardware capabilities, meaning that resource-intensive tools may only be available on more powerful subsystems, while lightweight variants are deployed on constrained edge devices.
\fwcomp{Subsystem \textit{n}} represents \underline{one} of $N$ subsystems in the computing architecture, e.g., a single edge device, where the \fwn instance communicates with other \fwn instances deployed on the rest of the architecture (i.e., the \fwcomp{Subsystems of Interest}).
With the term ``subsystem,'' we refer to a single independent computing node.
As part of \fwn, we use two types of MAPE-K loops: 
First, \fwblue{\fwcomp{Local MAPE-K}} represents the ``traditional'' adaptation loop commonly found in a self-adaptive system, running locally and only internally on each subsystem.
Second, the adaptation loop \fworange{\fwcomp{Global MAPE-K}} has a dedicated communication channel to other subsystems in the architecture, i.e., other edge and fog devices running a \fwn instance.

\fwblue{\fwcomp{Local MAPE-K}}: The \fwcomp{Local Runtime Monitor} gathers data from the \fwcomp{Managed system} and forwards it to the \fwcomp{Local Adaptation Middleware}.
The \fwcomp{Local SL Manager} (SL:~Security Level) plans an appropriate security level (cf.~\citesec{sec-levels}).
Finally, a \fwcomp{Local Execution Adapter} executes the planned adaptations on the \fwcomp{Managed system} ($\mapecharlocal{M}\Rightarrow\mapecharlocal{A}\Rightarrow\mapecharlocal{P}\Rightarrow\mapecharlocal{E}$).

\fworange{\fwcomp{Global MAPE-K}}: This adaptation loop serves two purposes: 
$(i)$ The loop allows for holistic adaptations together with other subsystems (cf. \textbf{AC~1}) and $(ii)$ ensures a prompt response in the event of an attack (cf.~\textbf{AC~3}).

$(i)$~The \fwcomp{Global SL Monitor} forwards adaptation information from the other \fwcomp{Subsystems of Interest}.
In our running case for a fog subsystem, the subsystems of interest are the superordinate server subsystems and the subordinate edge subsystems.
The adaptations, i.e., the security levels of other computing systems, are then forwarded to the \fwcomp{Local Adaptation Middleware}. 
The  \fwblue{\fwcomp{Local MAPE-K}} can then, besides the local monitoring data, also take comprehensive information from other connected systems into account for performing adaptations ($\mapecharapi{M}/\mapecharlocal{M}\Rightarrow\mapecharlocal{A}\Rightarrow\mapecharlocal{P}\Rightarrow\mapecharlocal{E}$).

$(ii)$ Security incidents can lead to isolated subsystems, disrupting their integration into the \fwcomp{Subsystems of Interest} (cf. SoS in \textbf{AC~2}).
When this happens, it is critical to ensure prompt adaptation times to prevent other subsystems from being compromised.
Disruptions can cause delays in the adaptation time, e.g., due to unanswered requests during an attack (cf.~\textbf{AC~3}).
The \fworange{\fwcomp{Global MAPE-K}} takes these disruptions into account by adapting the \fwcomp{Managed API} (Application Programming Interface). 
The API configuration uses two operational modes: Full Adaptation (FA) when \underline{all} other subsystems are available and Partial Adaptation (PA) when \underline{at least one} subsystem of interest is not reachable.
In both modes, the \fworange{\fwcomp{Global MAPE-K}} will be executed, and the \fwcomp{Managed API} adapts continuously.
The difference lies in the number of \fwcomp{Subsystems of Interest} checked for adaptation updates (cf. details in \citesec{global-adapt}).
The switch between these modes is decided within the \fworange{\fwcomp{Global MAPE-K}}:
The \fwcomp{Managed API} gathers the \textit{Global SL
Adaptations} and forwards the \textit{Network Status Data} to the \fwcomp{Global Network Monitor}.
The \fwcomp{Global Adaptation Middleware} checks the connections to other subsystems and detects if a subsystem in the \fwcomp{Subsystems of Interest} is unresponsive.
The \fwcomp{Global Config Manager} then triggers the respective mode, and the \fwcomp{Global Execution Adapter} carries out the adaptation by reconfiguring the \fwcomp{Managed API}
($\mapecharglobal{M}\Rightarrow\mapecharglobal{A}\Rightarrow\mapecharglobal{P}\Rightarrow\mapecharglobal{E}$).

\begin{figure*}[t!]
    \centering
    \includegraphics[width=\textwidth]{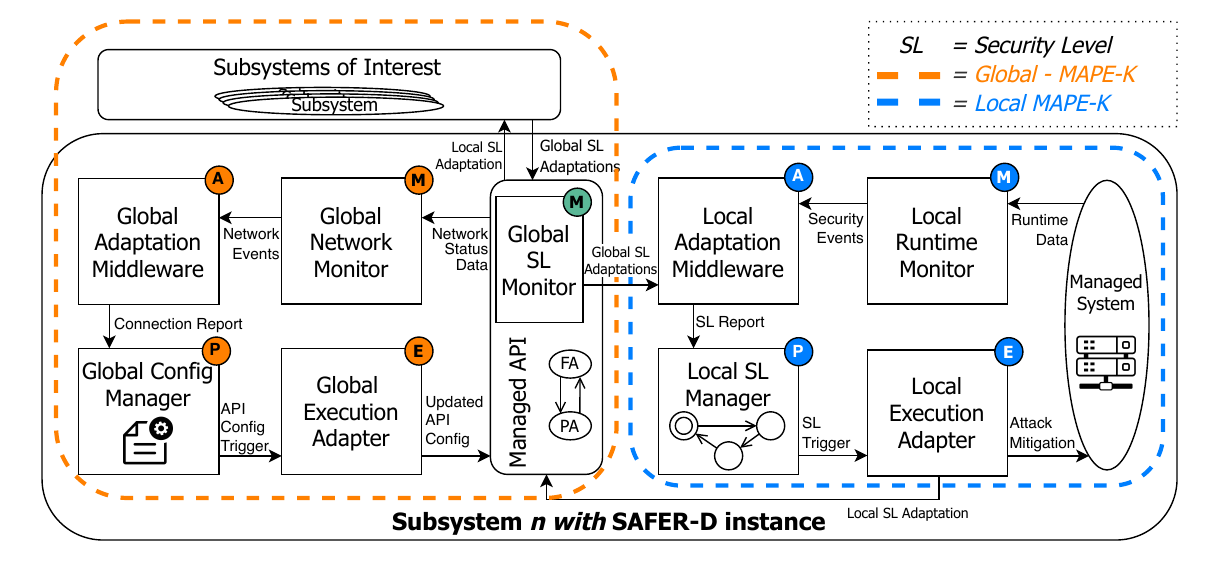}
    \caption{High-level Overview of \fwn.}
    \label{fig:high-level-fw}
\end{figure*}

\subsection{Security Levels}
\label{sec:sec-levels}

Security (criticality) levels (SLs) in \fwn represent varying degrees of protective measures specifically tailored to the current risk and criticality of a system (cf.~\textbf{AC~4}). 
The concept is inspired by so-called ``readiness levels,'' found, for instance, in the US military's DEFense readiness CONdition (DEFCON) stages~\cite{theisenDEFCONLevels2023}.
They enable dynamic adaptation of security mechanisms, ensuring system can escalate or de-escalate their defenses based on evolving threats or environmental conditions. 
We introduce SLs as part of  \fwn's \fwcomp{Local SL Manager} to adjust security based on dynamic changes (cf.~$x$ and/or $p$ in Eq.~\eqref{eq:risk}). 
For example, in our running case, under normal conditions, an edge device operates at DEFCON~5. 
If unusual behavior suggests a potential attack, it escalates to DEFCON~4, increasing the monitoring of system load, as sudden spikes may indicate a DDoS attack.
The DEFCON level continues to adapt as risks rise. 
The highest escalation is DEFCON~1 in critical cases (e.g., a confirmed DDoS attack) where the system may even shut down. 
The discrete SL design encourages security engineers to define clear, scenario-specific countermeasures per threshold, though we acknowledge this limits flexibility in multi-threat situations; addressing such conflicts will be a focus in future iterations of the framework.

The \fwcomp{Local Adaptation Middleware}, and the subsequent \fwcomp{Local SL Manager}, consider two sources when deciding on the appropriate SL.
The local \textit{Security Events} provided by the \fwcomp{Local Runtime Monitor} (\mapecharlocal{M}) and the \textit{Global SL Adaptations} of the other \fwcomp{Subsystems of Interest} in the architecture forwarded by the \fwcomp{Global SL Monitor} (\mapecharapi{M}).
SAFER-D supports a human-in-the-loop approach for returning to less restrictive security levels.

\subsection{Full and Partial Adaptation Mode}
\label{sec:global-adapt}


The \fworange{\fwcomp{Global MAPE-K}} leverages two distinct operational modes to speed up the global adaptation loop (cf. \citesec{core-components}): \textit{Full Adaptation (FA)} and \textit{Partial Adaptation (PA)}.
In both modes, \fwn tries to gather adaptation information from other \fwcomp{Subsystems of Interest}.

In a perfect world, a single subsystem can always communicate with the other subsystems of interest. 
Consider the example in \citefig{uc-1-arch}, showing a conceptual edge computing architecture with server, fog, and edge computing subsystems. 
For instance, \fwcomp{\#2} communicates updates to its superordinate server subsystem \fwcomp{\#1} and its subordinate subsystems \fwcomp{\#6} and \fwcomp{\#4}.
During FA, full adaptation within the DCA is possible, and every subsystem can receive respective adaptation updates from every other subsystem.
For the example in \citefig{uc-1-arch}, the subsystems adapting and exchanging data consists of all subsystems: $[$\fwcomp{\#1}, \fwcomp{\#2}, \fwcomp{\#3}, \fwcomp{\#4}, \fwcomp{\#6}, \fwcomp{\#7}, \fwcomp{\#5}, \fwcomp{\#8}$]$.

\begin{figure*}
     \centering
     \begin{subfigure}[b]{0.49\textwidth}
        \centering
        \includegraphics[width=\textwidth]{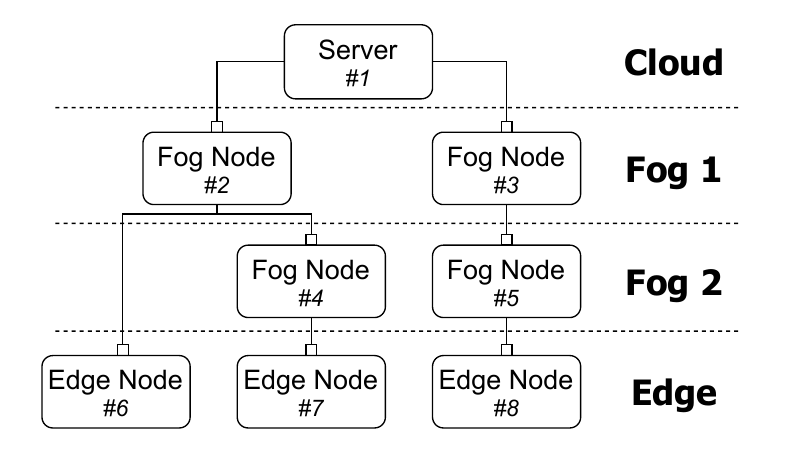}
        \caption{Architecture during \textit{FA}.}
        \label{fig:uc-1-arch}
     \end{subfigure}
     \hfill
     \begin{subfigure}[b]{0.49\textwidth}
         \centering
         \includegraphics[width=\textwidth]{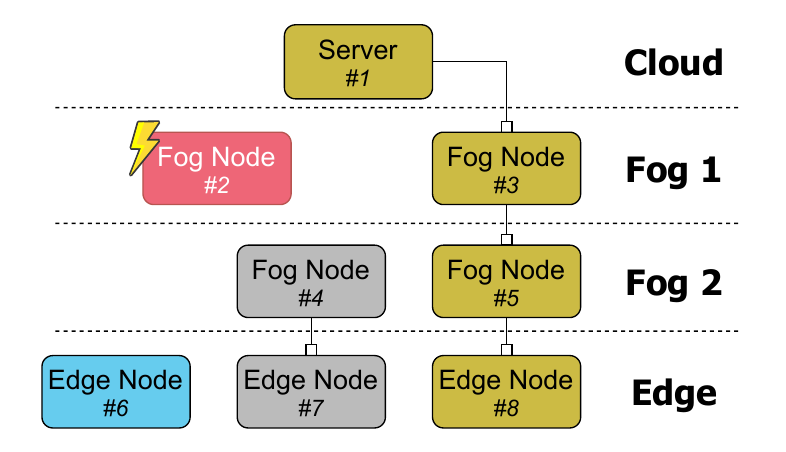}
         \caption{Successful attack at subsystem \texttt{\#2} and resulting \textit{PA} adaptation subgroups.} 
         \label{fig:uc-1-fba}
     \end{subfigure}
        \caption{Edge computing architecture in the use case scenario.}
        \label{fig:uc-1-archs}
\end{figure*}

In practice, a subsystem of interest can become unresponsive or unavailable due to an attack.
In such a case, fast and timely adaptation becomes one of the most important properties for preventing and potentially repelling attacks on other uncompromising subsystems.
The unavailable subsystem interferes with this goal. 
For instance, when subsystem $A$ sends a request to the unresponsive subsystem $B$.
$A$ cannot continue the adaptation until it reaches a timeout, thereby delaying the adaptation loop.
The problem is exacerbated when more than one subsystem is already compromised. -- In such a case, \fwn uses the  PA mode, displayed in \citefig{uc-1-fba}.
In the example, subsystem \textcolor{pinkish}{\fwcomp{\#2}} is unresponsive due to an attack.
As part of \fwn's \fworange{\fwcomp{Global MAPE-K}} loop, the remaining (still available) subsystems form adaptation subgroups based on the availability of connections.
In the example in \citefig{uc-1-fba}, the architecture is split into three adaptation subgroups: \textcolor{oliveish}{$[$\fwcomp{\#1}, \fwcomp{\#3}, \fwcomp{\#5}, \fwcomp{\#8}$]$}, \textcolor{greyish}{$[$\fwcomp{\#4}, \fwcomp{\#7}$]$}, and \textcolor{blueish}{$[$\fwcomp{\#6}$]$}.
Splitting the architecture and excluding the unresponsive subsystem \textcolor{pinkish}{\fwcomp{\#2}} helps maintain adequate adaptation times.
The splitting (i.e., the mode switch) in \fwn is performed by reconfiguring the \fwcomp{Managed API} and dictating which subsystems of the \fwcomp{Subsystems of Interest} should be requested for adaptation updates (i.e., contribute to the \textit{Global SL Adaptations}) and which are ignored.
As part of the \fworange{\fwcomp{Global MAPE-K}} loop, a recovery strategy is in place to check the responsiveness of unavailable subsystems, similar to a heartbeat function.
Once a subsystem is up and running again, the subsystem is reintegrated into the adaptation set until the system can switch back to the FA mode.


In the following, we further describe the \underline{interplay} of the two MAPE-K loops. \fwn is efficiently greedy in the sense of how it conducts adaptations.
The more subsystems are available, the more information can be used.
\fwn always tries to connect to the other  \fwcomp{Subsystems of Interest} as part of the \fworange{\fwcomp{Global MAPE-K}}.
When this is impossible, it still uses its PA mode to receive as many timely adaptations as possible.
In the worst case, no other subsystems are available (i.e., \mapecharapi{M} is exhausted).
In this case, \fwn can at least adapt locally (i.e., \mapecharlocal{M} is the only source for adaptations).
The interplay of the loops is also visualized in \citefig{flowchart}.
The novelty of \fwn lies in its flexibility in different situations.
\fwn is capable of dealing with interruptions and can, whenever necessary, adapt so that two MAPE-K loops are always running efficiently: One to adapt the \fwcomp{Managed System}, and one to adapt the \fwcomp{Managed API}.
This flexibility directly adheres to the resilience of the architecture \fwn is deployed to and marks the main contribution of our framework.

\begin{figure*}[t!]
    \centering
    \includegraphics[width=\textwidth]{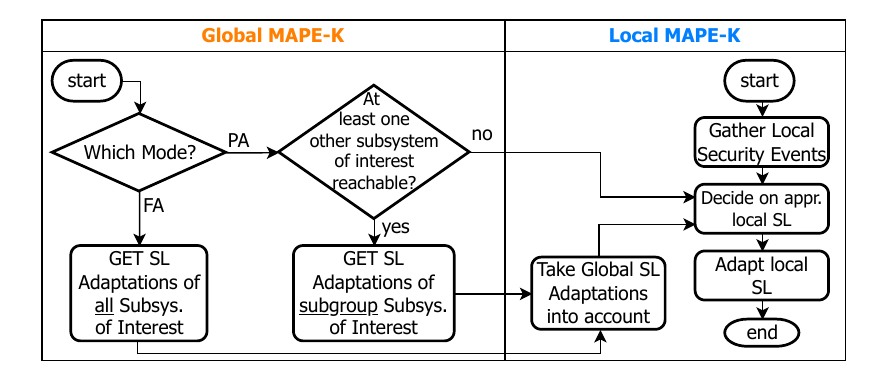}
    \caption{Simplified flowchart illustrating  \fworange{\fwcomp{Global MAPE-K}}'s influence on the \fwblue{\fwcomp{Local MAPE-K}}.}
    \label{fig:flowchart}
\end{figure*}

\section{Evaluation}
\label{sec:eval}

We executed a series of rigorous experiments. We evaluated  the \textit{Feasibility} and  \textit{Security} in an earlier framework version~\cite{stadlerCyberResilientEdgeComputing2024} here, we focus on \textit{Applicability} and \textit{Efficiency}. to evaluate our framework's general applicability and efficiency.
In this section, we first describe the research questions, use case scenarios, evaluation setup, results, and finally, the answers to our research questions. 

\RQ{\textit{(Applicability)} To what extent can \fwn be applied to execute security adaptations, and what is the integration effort when embedding it into existing system architectures?}\label{rq1}
With this first research question, we aim to qualitatively evaluate the applicability of \fwn. 
Applicability is essential to bridge theory and practice, and we try to approach an answer by addressing some of the key factors used in similar evaluation setups (see~\cite{leszczynaAimingMethodsWider2021}).
For measuring the integration effort, we report on the reusability of components and the required resources to use \fwn in a given scenario, i.e., the time and Lines of Code (LoC) it takes to configure and run \fwn.

\RQ{(Efficiency) To what extent can \fwn be used to execute security adaptations efficiently?}\label{rq2} With the second research question, we quantitatively determine the efficiency of \fwn.  
Since a timely adaptation is paramount during an attack (cf. \cite{coppolinoSelfadaptationbasedApproachResilience2023} and \textbf{AC~3}), short adaptation times are a crucial indicator of \fwn's efficiency. 
We assess the \textit{Time to Adapt} ($TtA$) on the deployed architecture of all MAPE-K loops.
First, we measure the \textit{Time to Adapt for Security Levels} ($TtA_{SL}$) to evaluate the adaptation times of $\mapecharapi{M}/\mapecharlocal{M}\Rightarrow\mapecharlocal{A}\Rightarrow\mapecharlocal{P}\Rightarrow\mapecharlocal{E}$.
Hence, the adaptations from one SL to another. 
Second, we measure the \textit{Time to Adapt for Global Adaptations} ($TtA_{G}$) to measure the switch between PA and FA to form adaptation subgroups.
Hence, $TtA_{G}$ reflects the time it takes for a $\mapecharglobal{M}\Rightarrow\mapecharglobal{A}\Rightarrow\mapecharglobal{P}\Rightarrow\mapecharglobal{E}$ loop configuration to take effect on the \fwcomp{Managed API}.

\subsection{Use Case Scenario}

Our use case is motivated by a real-world example provided by our industry partner \href{https://www.engelglobal.com/en/home}{ENGEL Austria GmbH}. 
The company is a large machine manufacturing enterprise operating in over 80 countries worldwide, with several thousand employees. The company is one of the leading manufacturers of industrial injection molding machines.
With its globalized status, the company provides multiple large edge computing architectures distributed worldwide for its customers.
Machine data (e.g., production cycles, usage data) is collected and distributed to central cloud computing nodes via a hierarchically structured edge computing architecture.
This data is used, among other things, for predictive maintenance, remote support, and data analysis to increase production efficiency and reduce energy consumption and scrap. 
Multiple machines are connected via edge devices, which are again connected to fog computing nodes and ultimately report to a central cloud server.
Although the machines can be operated standalone, downtime of the edge computing architecture due to a security incident or a cyberattack can impede normal operation and cause financial losses.
With its underlying edge computing architecture and security requirements, the company provides an excellent example for evaluating the different aspects of \fwn.
For the evaluation, we focused on DDoS attacks on the edge computing architecture~\cite{xiaoEdgeComputingSecurity2019}. More specifically, we used \fwn to adapt in the event of an Internet Control Message Protocol (ICMP) flood attack. 
In such an attack, an excessive number of ping requests are sent to the target, thereby clogging up its resources, such as network bandwidth and processing capabilities.
The use case's goal is to dictate security-level adaptations among the edge computing subsystems during an ICMP flood attack. 
If a system becomes unavailable due to the attack, the respective subsystems shall switch into the PA mode to allow for respective adaptations within the remaining adaptation subgroup.

\subsection{Evaluation Setup}

Based on the conceptual framework above, we created a prototype implementation in Python to conduct the evaluation.
In the following, we provide a brief overview of \fwn's instantiation (see details in suppl. material).
The industry partner provided us with edge devices (hardware) utilized on-site at their manufacturing plants.
We use these devices to replicate a typical edge computing architecture for the evaluation. 
The evaluated architecture is depicted in \citefig{uc-1-arch} and consists of four layers: one cloud, two fog, and one edge layer.%

\textbf{\RQref{1}:} The \fwcomp{Local Runtime Monitor}\footnote{
The specificity concerns the selection of monitored properties, not the underlying mechanism. Industrial systems often expose extensive metrics (e.g., \href{https://www.netdata.cloud/}{Netdata}), allowing the \fwcomp{Local Runtime Monitor} to function as a configurable filter or bridge
}, \fwcomp{Local Execution Adapter} \textit{(together 96 LoC)}, and the initial \textit{Architecture Configuration} (\textit{47 LoC}) are use-case-specific; the remaining implementation is reusable for other use cases (\textit{approx. 3h}).
We created three use-case-specific SLs with increasing measures for a potential ICMP flood attack. The levels are implemented using the Python state machine package~\cite{macedoFgmacedoPythonstatemachine2024} (\textit{70 LoC}, \textit{approx. 1.5h}) and are defined as follows: Level 3 -- Normal Readiness,  Regular monitoring; Level 2 -- Moderate Readiness, Rate limiting; Level 1 -- Maximum Readiness, Block entirely.
Each SL represents a state in the state machine. In the event of an attack, the \fwblue{\fwcomp{Local MAPE-K}} triggers state transitions from one state to another. The stepwise transitions are in place for control and dependency management.
SAFER-D fully supports configurable SL transitions (e.g., skipping a SL) when needed, which can be enabled by modifying the state machine definition.
At runtime, each subsystem periodically issues heartbeat requests to check for the SLs of the other connected subsystems.
For instance, \fwcomp{\#2} sends a heartbeat every ten seconds (interval aligns with typical machine cycle times, allowing heartbeats to be sent alongside operational data), requesting the SLs of \fwcomp{\#1}, \fwcomp{\#6}, and \fwcomp{\#4}. If one of the connected subsystems of interest responds with a higher criticality level than the one currently used for \fwcomp{\#2}, \fwcomp{\#2} adapts, i.e., transitions to the most critical SL. 
We prioritize the most restrictive SL to ensure rapid and effective response in high-impact scenarios like DoS attacks; in less time-critical contexts, incorporating human-in-the-loop decision-making can offer a more balanced trade-off between security and functionality.
Furthermore, RESTful communication allows us to easily identify whether a device is unresponsive: If a request times out, the prototype adapts accordingly using global adaptations.
The global adaptations are implemented by adding functionality to the periodic heartbeat checks.
Each system sends out a tree traversal (REST calls), checking which subsystems are still reachable.%

\textbf{\RQref{2}:} We evaluated $TtA_{SL}$ by monitoring our deployed framework operating on top of the edge computing architecture.
Adaptations are checked every ten seconds via a heartbeat (i.e., \fwblue{\fwcomp{Local MAPE-K}} and \fworange{\fwcomp{Global MAPE-K}} are periodically triggered).
 The goal is to capture the time from the beginning of an adaptation cycle (i.e., the start of the heartbeat) until an adaptation is detected (monitored), analyzed, processed, and executed. For instance, when considering \citefig{uc-1-arch}, \fwcomp{\#1} is before the adaptation heartbeat in \textit{Level 3}.
\fwcomp{\#2} is under attack and, therefore, in \textit{Level 2}.
For $TtA_{SL}$, we measure the time from the beginning of the heartbeat from \fwcomp{\#1} until \fwcomp{\#1} completes the adaptation to \textit{Level 2}. -- 
We chose \fwcomp{\#1} as the monitored subsystem to control the influence of network depth:
The closer the attacked subsystem is to the one monitored (in our case \fwcomp{\#1}), the faster it can adapt.
For instance, \fwcomp{\#2} is closer to \fwcomp{\#1} than \fwcomp{\#8}.
Since the heartbeat adaptation checks occur sequentially, network depth matters.
Therefore, \fwcomp{\#1} provides the best-case and worst-case scenario. It has systems connected directly (e.g., \fwcomp{\#2}) and, at the same time, exhibits the longest depth to traverse. 
During the evaluation, we seeded predefined SL triggers for each system depicted in \citefig{uc-1-arch}. We measured the $TtA_{SL}$ and validated that the adapted SLs were correct according to our seed, i.e., checking if \fwcomp{\#1} transitions to \textit{Level 2} when it is supposed to.
We repeated this process for every subsystem $N=100$ (i.e., a total of 700 SL adaptations).
We employed a similar procedure for $TtA_{G}$.
We randomly terminated components in \citefig{uc-1-arch}. The rest of the edge computing architecture had to respond accordingly and globally adapt to the PA mode.
We measured the time again from the start of a heartbeat cycle until the system transitioned to the PA mode (i.e., all adaptation subgroups were formed).
We chose \fwcomp{\#1} again as our monitored component to control for network depth. Every subsystem in \citefig{uc-1-arch} is randomly terminated $N=50$ times (350 in total).%

\subsection{Evaluation Results}

\textbf{\RQref{1}:} As a first step, we validated that every SL adaptation was carried out correctly as expected.
Each change in the SL of an attacked subsystem in the architecture resulted in the expected change of \fwcomp{\#1}. For the global adaptations, we can also confirm that \fwn's implementation executed correct adaptations in the architecture every time. 
Regarding the integration effort, we identified the components that would need re-implementation and counted the LoC and the time it took us to implement them.
A total of 213 LoC are use-case-specific, which took us approximately 4-5h.
Since every subsystem runs the same instance of \fwn, the development and configuration effort must be invested once.
The code is pulled via version control on every subsystem and is ready to run without further configuration.
Considering that the majority of \fwn's components can be reused for other use cases, 213 LoC and 4.5h of effort represent a considerably low effort with respect to the benefits \fwn can bring to such an architecture.%

\textbf{\RQref{2}:} \citefig{boxplots-uc-1-sec-lvl-adapt} and \citefig{boxplots-uc-1-global-adapt} show the boxplots of our quantitative analysis. 
One can notice the considerably long adaptation times for $TtA_{G}$, especially compared to $TtA_{SL}$. However, 3 seconds of the values in $TtA_{G}$ can be attributed to the initial HTTP timeout. An HTTP request always waits 3 seconds (the time an SL adaptation would take additionally) for a response during the heartbeat. After the waiting period, the system is deemed unresponsive. Therefore, these numbers always contain a fixed constant of 3000 ms.
The adaptation times for the $TtA_{SL}$ remain consistent across runs (cf. distribution in \citefig{boxplots-uc-1-sec-lvl-adapt}). The average median of $TtA_{SL}$ is 344.86 ms (roughly comparable to the 333 ms in~\cite{rieglerDistributedMAPEKFramework2023}). Therefore, most of $TtA_{SL}$ are considerably lower than half a second.
Similarly, for $TtA_{G}$, the average median adaptation time is 4543.29 ms, which means 1543.29 ms (total minus 3 seconds timeout) solely for the adaptation. However, adaptation times are still longer compared to $TtA_{SL}$. 
The reason is that once a subsystem is unresponsive, the system again traverses through the tree to determine the subgroup.
The traversal takes time, as shown in $TtA_{G}$.
The distribution of adaptation times among the subsystems can be considered equally stable for all subsystems for $TtA_{SL}$.
For $TtA_{G}$, the distribution is dense for \fwcomp{\#6}, \fwcomp{\#4}, \fwcomp{\#5}, and \fwcomp{\#8}.
\fwcomp{\#2} and \fwcomp{\#3} yielded a rather loose distribution, although their median is similar to the other subsystems.

\begin{figure*}[ht]
    \begin{minipage}[t]{\textwidth}
        \centering
        \setcounter{subfigure}{0}
        \begin{subfigure}[b]{0.49\textwidth}
            \centering
            \includegraphics[width=\textwidth]{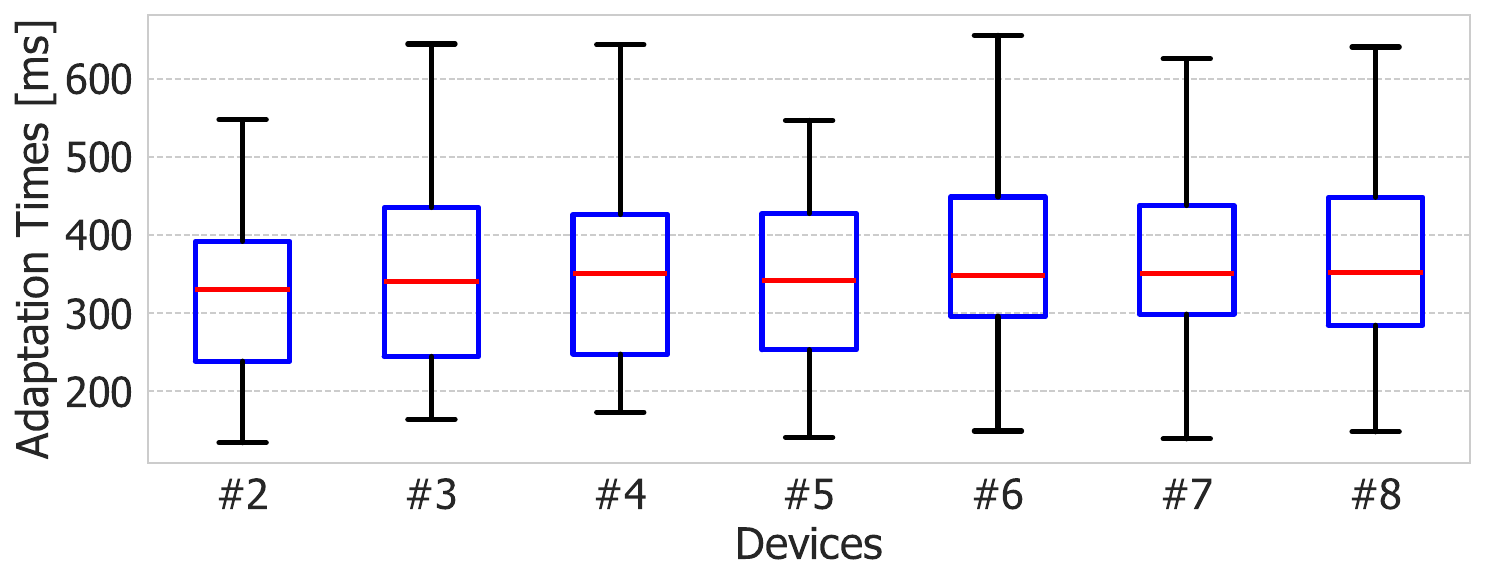}
            \caption{Boxplot for $TtA_{SL}$.}
            \label{fig:boxplots-uc-1-sec-lvl-adapt}
        \end{subfigure}
        \hfill
        \begin{subfigure}[b]{0.49\textwidth}
            \centering
            \includegraphics[width=\textwidth]{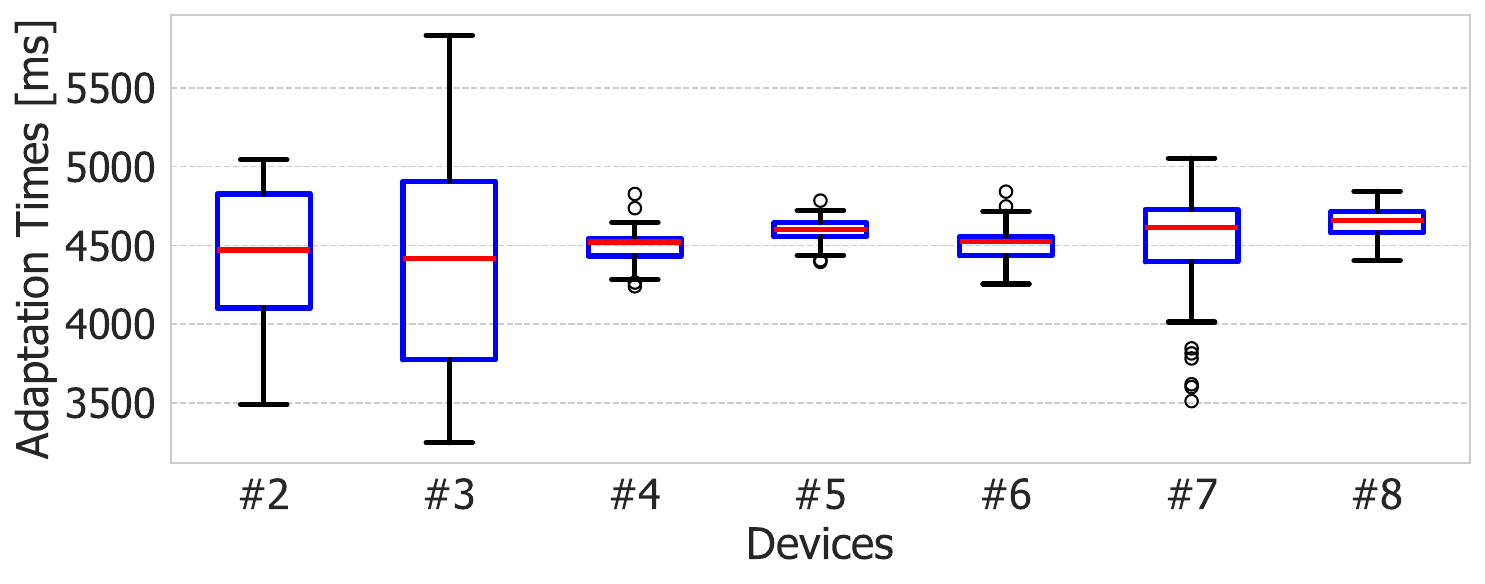}
            \caption{Boxplot for $TtA_{G}$.}
            \label{fig:boxplots-uc-1-global-adapt}
        \end{subfigure}
        \captionof{figure}{Boxplots of $TtA_{SL}$ \& $TtA_{G}$.}
        \label{fig:boxplots-eval}
    \end{minipage}\vspace{-1em}
\end{figure*}

\subsection{Answers to Research Questions}
\label{sec:discuss}

\textbf{Answer to \textbf{\RQref{1}}}:
The use case was inspired by a real-world industrial example. \fwn was successfully implemented within the edge computing architecture similar to the one utilized by the partnering company. 
This application validated the framework's capability to address practical challenges in an industry-relevant context.
Although not discussed here, due to space constraints, we deployed \fwn to a second use case focused on web authentication, further showcasing \fwn's adaptability to different domains and highlighting its ability to cater to security requirements in diverse settings (details can be found in the supplemental material; cf. data availability).
Both prototypical implementations are publicly available, enabling users to explore and leverage the adaptation mechanisms provided by \fwn.
The implementation effort for \fwn's adaptation components in our scenario offers a first indication of manageable integration (considering time and LoC), though further validation is needed. 
A total of 21h were invested to implement \fworange{\fwcomp{Global MAPE-K}} components and the \fwcomp{Local Adaptation Middleware} and \fwcomp{Local SL Manager} (i.e., the reusable parts of \fwn); the \fwcomp{Local Runtime Monitor} and \fwcomp{Local Execution Adapter} are use-case-specific and took us approx. 4.5h. 
We found the PA mode particularly helpful during development since even a network misconfiguration  immediately resulted in respective global adaptations. 
For this reason, we find \fwn's global adaptations helpful not only in the event of a security incident but also in maintenance or operational malfunctions.
In summary, through these applications, we demonstrated \fwn's applicability in real-world scenarios.
The initial results indicate that integration is feasible; however, a more comprehensive evaluation is required for general claims.

\textbf{Answer to \textbf{\RQref{2}}}:
We comprehensively evaluated \fwn's efficiency by quantitatively analyzing data from our use case, measuring adaptation times for SL adaptations ($TtA_{SL}$) and global adaptations ($TtA_{G}$).
Adaptation times were sufficiently efficient for the given scenarios.
Our analysis suggests that the centrality of the subsystem being considered influences adaptation times across the architecture (cf.~\citefig{boxplots-uc-1-global-adapt}). 
In general, top-level systems require longer adaptation times compared to low-level systems.
Optimized implementation strategies can improve performance, such as parallel tree traversal for adaptation checks.
While our findings highlight areas with potential for further optimization, the prototypical implementations provide a proof of concept. 
The results affirm that \fwn delivers promising efficiency, making it a viable framework for dynamic system adaptations in distributed computing systems.

%

\section{Threats to Validity}
\label{sec:threats}



Like any study, our work faces threats to validity. 
For conclusion validity, we used quantitative metrics $TtA_{SL}$ and $TtA_{G}$ and repeated runs to ensure consistency, though relying solely on time-based measures limits insight into security–functionality trade-offs; broader quantitative applicability metrics could offer a more comprehensive picture.
Regarding internal validity, we controlled the experimental setup to isolate the framework’s effect on adaptation times, minimizing the influence of hardware / network factors, although real-world deployments may introduce unforeseen variables. 
For external validity, our use cases and metrics (e.g., time, LoC) serve as a foundation for generalization; however, further validation in diverse environments is necessary to confirm broader applicability.


\section{Related Work}
\label{sec:rel-work}

\textbf{Multi-level Adaptation}:
In their work, Jahan~\etal~\cite{Jahan_etal_2020} propose a framework for dynamically maintaining functional and security concerns in autonomous systems, ensuring coordination between multiple MAPE-K feedback control loops. An additional MAPE-SAC loop is introduced that emphasizes security-related adaptations. Similarly, also employing a multi-feedback loop approach,  Vromant~\etal~\cite{vromant2011interacting} relied on intra-loop and inter-loop coordination of multiple MAPE-K loops to perform coordinated adaptation actions.
Braberman~\etal~\cite{brabermanMORPHReferenceArchitecture2015} present MORPH, a reference architecture for self-adaptation based on the MAPE-K loop.
MORPH consists of three layers for goal management, strategy management, and strategy enactment with different reconfiguration strategies.
Ben Halima~\etal~\cite{benhalimaMAPEKPatternsSelfadaptation2023} introduce a set of MAPE-K design patterns tailored for decentralized control in self-adaptive CPSs.
Gerostathopoulos~\etal~\cite{gerostathopoulosSelfadaptationSoftwareintensiveCyber2016} propose IRM-SA, an Invariant Refinement Method for Self-Adaptation, tailored to ensure dependability and adaptivity in software-intensive CPSs. 

\textbf{Security Adaptation}:
Fotohi~\etal~\cite{fotohiAgentbasedSelfprotectiveMethod2020} propose an Agent-based Self-Protective method (ASP-UAVN) inspired by the human immune system to enhance secure communication in Unmanned Aerial Vehicle Networks. 
Riegler~\etal~\cite{rieglerDistributedMAPEKFramework2023} introduce DSEC4IoT, a distributed MAPE-K framework for self-protective IoT devices, enabling local and centralized monitoring, analysis, planning, and execution of security measures.
Jones~\etal~\cite{jonesDefeatingDenialofserviceAttacks2019} present Crispy, a CRISPR-inspired (bacterial adaptive immune system) resiliency mechanism to protect N-variant systems from DoS attacks by leveraging automatic attack signature generation.
Finally, Skandylas~\cite{skandylasDesignAnalysisSelfprotection2021} presents an approach for enhancing adaptive security in software-intensive systems by equipping them with self-protective capabilities, including runtime threat modeling, proactive adaptation, and decentralized trust-based mechanisms.

%
\begin{table*}[h]
    
\centering
\begin{tabular}{| c || c | c | c | c | c || c | c | c | c || c |}
        
            \hline
            & \cite{Jahan_etal_2020} & \cite{vromant2011interacting} & \cite{brabermanMORPHReferenceArchitecture2015} & \cite{benhalimaMAPEKPatternsSelfadaptation2023} & \cite{gerostathopoulosSelfadaptationSoftwareintensiveCyber2016} & \cite{fotohiAgentbasedSelfprotectiveMethod2020} & \cite{rieglerDistributedMAPEKFramework2023} & \cite{jonesDefeatingDenialofserviceAttacks2019} & \cite{skandylasDesignAnalysisSelfprotection2021} & \fwn \\
            \hline
 \textit{Multi-System}                 & \hball{Half} & \hball{Half} & \hball{None} & \hball{Full} & \hball{Full} & \hball{Half} & \hball{Full} & \hball{    } & \hball{Full} & \hball{Full} \\ \hline
 \textit{Security Focus}               & \hball{Full} & \hball{    } & \hball{    } & \hball{None} & \hball{    } & \hball{Full} & \hball{Full} & \hball{Full} & \hball{Full} & \hball{Full} \\ \hline
 \textit{Failure Resilience}           & \hball{    } & \hball{Full} & \hball{Half} & \hball{None} & \hball{Full} & \hball{Half} & \hball{Full} & \hball{Full} & \hball{    } & \hball{Full} \\ \hline
 \textit{Loop Interaction}             & \hball{Full} & \hball{Full} & \hball{Full} & \hball{Full} & \hball{Full} & \hball{Half} & \hball{Half} & \hball{None} & \hball{    } & \hball{Full} \\ \hline
 \textit{Decentral Coordination}       & \hball{    } & \hball{Half} & \hball{    } & \hball{Full} & \hball{Full} & \hball{Half} & \hball{None} & \hball{    } & \hball{Full} & \hball{Full} \\ \hline
\end{tabular}
\caption{Overview of Related Work; [\hspace{5pt}]=Aspect absent, [\hball{None}]=Aspect mentioned, [\hball{Half}]=Aspect partially addressed, [\hball{Full}]=Aspect fully addressed}
\label{tab:rel-work}
\end{table*}




We provide an overview of the addressed contents of related work and how it compares to \fwn in \citetable{rel-work}.
The table reveals that \fwn is closely related to Fotohi~\etal~\cite{fotohiAgentbasedSelfprotectiveMethod2020} and Riegler~\etal~\cite{rieglerDistributedMAPEKFramework2023}. -- 
Fotohi~\etal~\cite{fotohiAgentbasedSelfprotectiveMethod2020} focus on \textit{securing communication} between UAVs, relying on detecting network-layer attacks and isolating malicious nodes.
In contrast, \fwn \textit{protects the devices themselves} through hierarchical coordination (vs. purely peer-to-peer) and adaptive security levels (vs. strict isolation), enabling graded and context-aware responses.
Unlike Riegler~\etal~\cite{rieglerDistributedMAPEKFramework2023}, who rely on a central ``Managing Server'' to coordinate security adaptation across independently operating devices, \fwn enables fully decentralized coordination among autonomous subsystems.
Moreover, DSec4IoT supports a fixed two-level structure (server and devices) while \fwn introduces a multi-level hierarchy where adaptation decisions can propagate and adjust across layers.
Therefore, in case of connection loss, \fwn can adapt within subgroups while DSec4IoT adapts only locally.


\section{Conclusion}
\label{sec:conclusion}

In this paper, we presented \fwn, a novel self-adaptive security framework for DCAs. 
\fwn integrates diverse adaptation strategies to enable security adaptations, even in the event of system failures caused by attacks.
Our evaluation using a realistic use case scenario has confirmed that \fwn can be used in practice and that the adaptations are efficiently carried out.
As part of our ongoing and future work, we aim to extend our adaptation strategies, improve performance and scalability for large-scale architectures, and incorporate advanced runtime threat modeling techniques.

\subsection*{Data Availability}

Supplemental material on GitHub: \href{https://github.com/jku-lit-scsl/ecsa25_safer-d_uc-1}{Edge computing use case} / \href{https://github.com/jku-lit-scsl/ecsa25_safer-d_uc-2}{WebAuthn use case}.

\balance
\bibliographystyle{abbrv}
\bibliography{main}

\end{document}